\documentclass[letterpaper, 10pt, conference]{ieeeconf}
\IEEEoverridecommandlockouts
\usepackage{graphicx, bm, mathrsfs}
\usepackage{cite,array}
\usepackage{amssymb, dsfont, color}
\usepackage{epstopdf}
\usepackage[cmex10]{amsmath}
\usepackage{enumerate}
\usepackage{multirow}
\usepackage{booktabs}
\usepackage{threeparttable}
\usepackage{subeqnarray}
\usepackage{cases, stmaryrd}
\usepackage{stackengine}

\newtheorem{assumption}{Assumption}
\newtheorem{theorem}{Theorem}

\newtheorem{definition}{Definition}
\newtheorem{remark}{Remark}

\definecolor{bluencs}{rgb}{0.0, 0.53, 0.74}

\newcommand{\p}{\textrm{p}}
\newcommand{\rf}{\textrm{r}}
\newcommand{\ct}{\textrm{c}}
\newcommand{\f}{\textrm{f}}

\newcommand{\ax}{\textrm{a}}
\newcommand{\dom}{\textrm{dom}}
\DeclareMathOperator*{\esssup}{\textrm{ess.\,sup}}

\newcommand{\diag}{\textrm{diag}}

\newcommand{\tran}{\mathrm{T}}
\newcommand{\overbar}[1]{\mkern 1.5mu\overline{\mkern-1.5mu#1\mkern-1.5mu}\mkern 1.5mu}

\begin{document}

\title{\LARGE \bf Event-Triggered Tracking Control of Networked and Quantized Control Systems
\thanks{This work was supported by the H2020 ERC Consolidator Grant L2C, the Walloon Region and the Innoviris Foundation, the H2020 ERC Consolidator Grant LEAFHOUND, the Swedish Foundation for Strategic Research (SSF), the Swedish Research Council (VR) and the Knut och Alice Wallenberg Foundation (KAW).}
}

\author{Wei~Ren, Dimos V. Dimarogonas, and Rapha\"el Jungers
\thanks{W. Ren and R. Jungers are with ICTEAM institute, UCLouvain, 1348 Louvain-la-Neuve, Belgium. D. V. Dimarogonas is with Division of Decision and Control Systems, EECS, KTH Royal Institute of Technology, SE-10044, Stockholm, Sweden.
Email: \texttt{\small gtppwe@gmail.com, raphael.jungers@uclouvain.be, dimos@kth.se}.}
}

\maketitle

\begin{abstract}
This paper studies the tracking control problem of networked and quantized control systems under both multiple networks and event-triggered mechanisms. Multiple networks are to connect the plant and reference system with decentralized controllers to guarantee their information transmission, whereas event-triggered mechanisms are to reduce the information transmission via multiple networks. In this paper, all networks are independent and asynchronous and have local event-triggered mechanisms, which are based on local measurements and determine whether the local measurements need to be transmitted. We first implement an emulation-based approach to develop a novel hybrid model for tracking control of networked and quantized control systems. Next, sufficient conditions are derived and decentralized event-triggered mechanisms are designed to ensure the tracking performance. Finally, a numerical example is given to illustrate the obtained results.
\end{abstract}

\section{Introduction}
\label{sec-introduction}

The introduction of wired/wireless networks to connect multiple smart devices leads to networked control systems (NCS), the area of which includes three activities \cite{Zampieri2008trends}: control of networks; control over networks; and multi-agent systems. The presence of networks improves efficiency and flexibility of integrated applications, and reduces installation and maintenance time and costs \cite{Gupta2010networked, Heemels2010networked, Dolk2016output}. Different smart devices may be physically distributed and interconnected such that their communications are via different types of networks, which in turn result in many issues, such as transmission delays, packet dropouts, quantization, etc. Therefore, the main challenge is how to design the control scheme to limit the effects of the aforementioned network-induced issues and to achieve the desired performances while keeping the information transmission as minimal as possible. One suitable approach in this context is periodic event-triggered control (PETC) \cite{Heemels2012periodic, Garcia2016periodic, Yu2018explicit}, combining time-triggered control (TTC) \cite{Poveda2019hybrid, Gao2011sampled} and event-triggered control (ETC) \cite{Tabuada2007event, Mazo2011decentralized, Dimarogonas2011distributed}. The PETC allows the triggering condition to be evaluated with a predefined sampling period to decide the information transition, and leads to a balance between TTC and ETC by avoiding the continuous evaluation of the triggering condition \cite{Heemels2012periodic, Yu2018explicit}.

Many existing results on NCS focus mainly on stability analysis and stabilization control, and both TTC and ETC/PETC have been addressed \cite{Heemels2010networked, Nesic2004input, Heemels2012periodic, Dolk2016output}. However, tracking control, as a fundamental problem in control theory \cite{Ren2019tracking, Postoyan2014tracking}, received less attention \cite{Hong2006tracking, Cheng2016event, Postoyan2015event}. The main objective of the tracking control is to design controllers such that the plant can track the given reference trajectory as close as possible \cite{Tallapragada2013event, Biemond2013tracking}. In the tracking control, the controller consists of two parts \cite{Van2010tracking}: the feedforward part to induce the reference trajectory for the plant, and the feedback part to drive the plant to converge to the reference trajectory. As opposed to the traditional tracking control, the main challenge of the tracking control of NCS is that only local/partial information is transmitted to the plant due to limited-capacity communication networks. In addition, the information transmission via networks may be a error source affecting the tracking performance \cite{Van2010tracking}. Therefore, both network-induced errors and local interaction rules need to be considered simultaneously, and thus result in additional difficulties in the tracking performance analysis.

In this paper, we study the event-triggered tracking control problem for networked and quantized control systems (NQCS), where several issues caused by the network and quantization are included \cite{Nesic2009unified}. To this end, we implement an emulation-like approach as in \cite{Ren2019tracking, Postoyan2014tracking, Heemels2010networked}, and develop a novel hybrid model using the formalism in \cite{Cai2009characterizations} to address the event-triggered tracking control for NQCS, which is our first contribution. In particular, a general scenario is considered: multiple independent and asynchronous networks are applied to ensure the communication among different components. This scenario stems from many physical systems, where different communication channels are applied to connect sensors, controllers and actuators. Hence, this setting allows to recover the architectures in \cite{Ren2019tracking, Postoyan2014tracking, Van2010tracking} for NCS and \cite{Cheng2016event, Postoyan2015event, Ren2020event} for MAS as special cases. In particular, external disturbances are considered in \cite{Ren2020event}, whereas quantization effects are studied here. Based on this setting, a general hybrid model is developed to incorporate all the issues caused by multiple networks and decentralized event-triggered mechanisms (ETMs). Our second contribution is to apply the Lyapunov-based approach to investigate the effects of these issues on the tracking performance. Specifically, motivated by multiple Lyapunov functions approach, some reasonable assumptions are provided, the decentralized ETMs are designed to reduce the frequency of the information transmission, and the tradeoff between the maximally allowable sampling period (MASP) and the maximally allowable delay (MAD) is derived to guarantee the convergence of the tracking error with respect to the network-induced errors.

Preliminaries are presented in Section \ref{sec-preliminaries}. The tracking problem is formulated in Section \ref{sec-problemformation}, and a unified hybrid model is developed in Section \ref{sec-developandreformulate}. Lyapunov-based conditions and decentralized ETMs are derived in Section \ref{sec-mainresults}. A numerical example is presented in Section \ref{sec-illustration}. Conclusions and further research are stated in Section \ref{sec-conclusion}.

\section{Preliminaries}
\label{sec-preliminaries}

$\mathbb{R}:=(-\infty, +\infty)$; $\mathbb{R}_{\geq0}:=[0, +\infty)$; $\mathbb{R}_{>0}:=(0, +\infty)$; $\mathbb{N}:=\{0, 1, 2, \ldots\}$; $\mathbb{N}_{+}:=\{1, 2, \ldots\}$. Given two sets $\mathcal{A}$ and $\mathcal{B}$, $\mathcal{B}\backslash\mathcal{A}:=\{x: x\in\mathcal{B}, x\notin\mathcal{A}\}$. $|\cdot|$ denotes the Euclidean norm. Given two vectors $x, y\in\mathbb{R}^{n}$, $(x, y):=(x^{\tran}, y^{\tran})^{\tran}$ for simplicity of notation, and $\langle x, y\rangle$ denotes the usual inner product. $\mathds{E}$ denotes the vector with all the components being 1, $I$ denotes the identity matrix of appropriate dimension, and $\diag\{A, B\}$ denotes the block diagonal matrix made of the matrices $A$ and $B$. Given a function $f: \mathbb{R}_{\geq0}\rightarrow\mathbb{R}^{n}$, $f(t^{+}):=\limsup_{s\rightarrow0^{+}}f(t+s)$. A function $\alpha: \mathbb{R}_{\geq0}\rightarrow\mathbb{R}_{\geq0}$ is of class $\mathcal{K}$ if it is continuous, $\alpha(0)=0$, and strictly increasing; it is of class $\mathcal{K}_{\infty}$ if it is of class $\mathcal{K}$ and unbounded. $\beta: \mathbb{R}^{2}_{\geq0}\rightarrow\mathbb{R}_{\geq0}$ is of class $\mathcal{KL}$ if $\beta(s, t)\in\mathcal{K}$ for fixed $t\geq0$ and $\beta(s, t)$ decreases to zero as $t\rightarrow0$ for fixed $s\geq0$. A function $\beta:\mathbb{R}^{3}_{\geq0}\rightarrow\mathbb{R}_{\geq0}$ is of class $\mathcal{KLL}$ if $\beta(r, s, t)\in\mathcal{KL}$ for fixed $s\geq0$ and $\beta(r, s, t)\in\mathcal{KL}$ for fixed $t\geq0$.

Consider the hybrid system \cite{Cai2009characterizations}:
\begin{equation}
\label{eqn-1}
\begin{cases}
\dot{x}=F(x, w), \quad (x, w)\in C,  \\
x^{+}=G(x, w), \quad (x, w)\in D,
\end{cases}
\end{equation}
where $x\in\mathbb{R}^{n}$ is the system state, $w\in\mathbb{R}^{m}$ is the external input, $F: C\rightarrow\mathbb{R}^{n}$ is the flow map, $G: D\rightarrow\mathbb{R}^{m}$ is the jump map, $C$ is the flow set and $D$ is the jump set. For the hybrid system \eqref{eqn-1}, the following basic assumptions are presented: the sets $C, D\subset\mathbb{R}^{n}\times\mathbb{R}^{m}$ are closed; $F$ is continuous on $C$; and $G$ is continuous on $D$. In \eqref{eqn-1}, $x\in\mathbb{R}^{n}$ is defined on \textit{hybrid time domain}, which is denoted by $\dom x\subset\mathbb{R}_{\geq0}\times\mathbb{N}$ with the following structure: for each $(T, J)\in\dom x$, $\dom x\cap([0, T]\times\{0, \ldots, J\})$ can be written as $\bigcup_{0\leq j\leq J-1}([t_{j}, t_{j+1}], j)$ for some finite sequence of times $0=t_{0}\leq t_{1}\leq\ldots\leq t_{J}=T$. $(t', j')\preceq(t, j)$ if $t'+j'\leq t+j$. A solution $(x, w)$ to \eqref{eqn-1} is a function on the hybrid time domain satisfying the dynamics in \eqref{eqn-1} with the following property: $\dom x=\dom w$; $x(\cdot, j)$ with fixed $j$ is absolutely continuous; and $w(\cdot, j)$ with fixed $j$ is Lebesgue measurable and locally essentially bounded. A solution $(x, w)$ is \emph{maximal} if it cannot be extended. Define $\|w\|_{(t, j)}:=\max\left\{\esssup\limits_{(t', j')\in\dom w\setminus\Xi(w), (0, 0)\preceq(t', j')\preceq(t, j)}|w(t', j')|,\right.$ $\left.\sup\limits_{(t, j)\in\Xi(w), (0, 0)\preceq(t', j')\preceq(t, j)}\sup|w(t', j')|\right\}$, and $\|w\|:=$ $\sup_{(t, j)\in\dom w}\|w\|_{(t, j)}$, where $\Xi(w):=\{(t, j)\in\dom w: (t, j+1)\in\dom w\}$. $\mathfrak{S}_{w}(x_{0})$ is the set of all the maximal solutions to \eqref{eqn-1} with $x_{0}=x(0, 0)\in C\cup D$ and finite $\|w\|$.

\begin{definition}[\cite{Cai2009characterizations}]
The hybrid system \eqref{eqn-1} is \emph{input-to-state stable (ISS) } from $w$ to $x$, if there exist $\beta\in\mathcal{KLL}$ and $\gamma\in\mathcal{K}_{\infty}$ such that $|x(t, j)|\leq\beta(|x(0, 0)|,t, j)+\gamma(\|w\|_{(t, j)})$ for all $(t, j)\in\dom x$ and all $(x, w)\in\mathfrak{S}_{w}(x_{0})$.
\end{definition}

\section{Problem Formulation}
\label{sec-problemformation}

In this section, we first state the tracking control problem for the NQCS studied in this paper, and then present the detailed information transmission among the plant, the reference and the controller via multiple networks.

\subsection{Tracking Problem of NQCS}
\label{subsec-trackingproblem}

Consider the following nonlinear system
\begin{align}
\label{eqn-2}
\dot{x}_{\p}=f_{\p}(x_{\p}, u),\quad y_{\p}=g_{\p}(x_{\p}),
\end{align}
where $x_{\p}\in\mathbb{R}^{n_{\p}}$ is the system state, $u\in\mathbb{R}^{n_{u}}$ is the control input, and $y_{\p}\in\mathbb{R}^{n_{y}}$ is the plant output. Similar to \cite{Ren2019tracking, Postoyan2014tracking, Hong2006tracking, Cheng2016event, Postoyan2015event, Tallapragada2013event, Biemond2013tracking}, the reference system tracked by \eqref{eqn-2} is of the form:
\begin{align}
\label{eqn-3}
\dot{x}_{\rf}=f_{\p}(x_{\rf}, u_{\f}),\quad y_{\rf}=g_{\p}(x_{\rf}).
\end{align}
where $x_{\rf}\in\mathbb{R}^{n_{\rf}}$ is the reference state ($n_{\rf}=n_{\p}$), $u_{\f}\in\mathbb{R}^{n_{u}}$ is the feedforward control input, and $y_{\rf}\in\mathbb{R}^{n_{y}}$ is the reference output ($n_{y_{\rf}}=n_{y_{\p}}$). Assume that the reference system \eqref{eqn-3} has a unique solution for any initial condition and any input.

To track the reference system, the controller designed for \eqref{eqn-2} in the absence of the network and quantizer is $u=u_{\ct}+u_{\f}$, where $u_{\ct}\in\mathbb{R}^{n_{u}}$ is the feedback item from the following nonlinear feedback controller
\begin{align}
\label{eqn-4}
\dot{x}_{\ct}=f_{\ct}(x_{\ct}, y_{\p}, y_{\rf}),\quad u_{\ct}=g_{\ct}(x_{\ct}),
\end{align}
where $x_{\ct}\in\mathbb{R}^{n_{\ct}}$ is the state of the feedback controller; $u_{\f}\in\mathbb{R}^{n_{u}}$ is the feedforward item and is related to plant state and reference state \cite{Van2010tracking}. We assume that $f_{\p}$ and $f_{\ct}$ are continuous; $g_{\p}$ and $g_{\ct}$ are continuously differentiable.

Since the emulation-based approach is applied \cite{Heemels2010networked, Ren2019tracking, Postoyan2014tracking}, the feedback controller \eqref{eqn-4} is assumed to be designed for the network-free and quantization-free case. Hence, the objective of this paper is to implement the designed controller over both ETMs and multiple networks and quantizers, and to ensure that the assumed tracking performance of the system \eqref{eqn-2}-\eqref{eqn-4} will be preserved for the NQCS under reasonable assumptions and the designed decentralized ETMs.

\subsection{Information Transmission over Multiple Networks}
\label{subsec-informationtransmission}

The information is sampled via the sensors, quantized and then determined (by the ETM to be designed) to be transmitted via the network. Since the sensors and actuators may be of different types, the connection among the plant, the reference and the controller may be via multiple networks (e.g., wired/wireless networks \cite{Gupta2010networked, Dolk2016output}). Therefore, the information is transmitted via multiple networks, which are assumed to satisfy the following assumption.

\begin{assumption}
\label{asn-1}
In the case that the ETM is implemented, all the sensors and actuators are connected via $N\in\mathbb{N}_{+}$ independent and asynchronous networks.
\end{assumption}

For each network $i\in\mathcal{N}:=\{1, \ldots, N\}$, the information to be transmitted is denoted by $z_{i}:=(y^{i}_{\p}, y^{i}_{\rf}, u^{i}_{\f}, u^{i}_{\ct})\in\mathbb{R}^{n^{i}_{z}}$ with $n^{i}_{z}:=2n^{i}_{y}+2n^{i}_{u}$. The dynamics of $z_{i}$ is written as
\begin{align}
\label{eqn-5}
\dot{z}_{i}&=f^{i}_{z}(z_{i}, x_{\p}, x_{\rf}, x^{i}_{\ct}),
\end{align}
where $f^{i}_{z}$ can be computed explicitly via \eqref{eqn-2}-\eqref{eqn-4}. The dependence of $\dot{z}_{i}$ on $x_{\p}$ and $x_{\rf}$ comes from the potential dependence of $y^{i}_{\p}$ (or $y^{i}_{\rf}$) on $x_{\p}$ (or $x_{\rf}$). Denote $z:=(z_{1}, \ldots, z_{N})\in\mathbb{R}^{n_{z}}$ with $n_{z}:=\sum^{N}_{i=1}n^{i}_{z}$, and $\dot{z}=f_{z}:=(f^{1}_{z}, \ldots, f^{N}_{z})\in\mathbb{R}^{n_{z}}$. Because of the band-limited capacity of each network and spatial locations of its sensors and actuators, all the sensors and actuators of each network are grouped into $\ell_{i}\in\mathbb{N}_{+}$ nodes to access to the network, where $i\in\mathcal{N}$ \cite{Walsh2002stability, Nesic2004input}. Correspondingly, $z_{i}$ is partitioned into $\ell_{i}$ parts. For the $i$-th network, its sampling time sequence is given by $\{t^{i}_{j}: i\in\mathcal{N}, j\in\mathbb{N}_{+}\}$, which is strictly increasing. At $t^{i}_{j}$, one and only one node is allowed to access to the $i$-th network, and this node is chosen by an time-scheduling protocol; see Subsection \ref{subsec-protocol}. For the $i$-th network, the sampling intervals are defined as $h^{i}_{j}:=t^{i}_{j+1}-t^{i}_{j}$, where $i\in\mathcal{N}$ and $j\in\mathbb{N}_{+}$. Since it takes time to compute and transmit the information, there exist transmission delays $\tau^{i}_{j}\geq0$ such that the information is received at the arrival times $r^{i}_{j}=t^{i}_{j}+\tau^{i}_{j}$.

\begin{assumption}
\label{asn-2}
For the $i$-th network, $i\in\mathcal{N}$, there exist constants $T_{i}\geq\Delta_{i}\geq0$ and $\varepsilon_{i}\in(0, T_{i})$ such that $\varepsilon_{i}\leq h^{i}_{j}\leq T_{i}$ and $0\leq\tau^{i}_{j}\leq\min\{\Delta_{i}, h^{i}_{j}\}$ hold for all $j\in\mathbb{N}_{+}$.
\end{assumption}

In Assumption \ref{asn-2}, $T_{i}>0$ is called the \textit{maximally allowable sampling period (MASP)} for the $i$-th network, $\Delta_{i}\geq0$ is called the \textit{maximally allowable delay (MAD)}, and $\varepsilon_{i}>0$ is the minimal interval of two successive transmissions. The constant $\varepsilon_{i}$ is determined by hardware constraints \cite{Heemels2010networked}, and ensures the exclusion of Zeno phenomena. In the network-free case \cite{Yu2018explicit}, $\varepsilon_{i}\equiv0$ and $0<h^{i}_{j}\leq T_{i}$ in Assumption \ref{asn-2}. Note that the MASP and MAD are design parameters and  will be upper bounded in Subsection \ref{subsec-trackanalyz}.

The sampled information is quantized before being transmitted. For each network, each node $j\in\{1, \ldots, \ell_{i}\}$ has a quantizer. The quantizer is a piecewise continuous function $\bar{q}^{i}_{j}: \mathbb{R}^{n^{i}_{j}}\rightarrow\mathcal{Q}^{i}_{j}\subset\mathbb{R}^{n^{i}_{j}}$, where $\mathcal{Q}^{i}_{j}$ is finite. The following assumption is made for the quantizer; see also \cite{Liberzon2007input}.

\begin{assumption}
\label{asn-3}
For all $i\in\mathcal{N}$ and $j\in\{1, \ldots, \ell_{i}\}$, there exist $\mathfrak{m}^{i}_{j}>\mathfrak{n}^{i}_{j}>0$ and $\mathfrak{n}^{i}_{0j}>0$ such that for all $z^{i}_{j}\in\mathbb{R}^{n^{i}_{j}}$: i) $|z^{i}_{j}|\leq\mathfrak{m}^{i}_{j}\Rightarrow |\bar{q}^{i}_{j}(z^{i}_{j})-z^{i}_{j}|\leq\mathfrak{n}^{i}_{j}$; ii) $|z^{i}_{j}|>\mathfrak{m}^{i}_{j}\Rightarrow|\bar{q}_{j}(z^{i}_{j})|>\mathfrak{m}^{i}_{j}-\mathfrak{n}^{i}_{j}$; iii) $|z^{i}_{j}|\leq\mathfrak{n}^{i}_{0j}\Rightarrow\bar{q}^{i}_{j}(z^{i}_{j})\equiv0$.
\end{assumption}

In Assumption \ref{asn-3}, $\epsilon^{i}_{j}:=\bar{q}^{i}_{j}(z_{j})-z^{i}_{j}$ is defined as the quantization error. $\mathfrak{m}^{i}_{j}$ is the range of the $j$-th quantizer in $i$-th network, $\mathfrak{n}^{i}_{j}$ is the bound on the quantization error. The condition i) gives a bound on the quantization error when the quantizer does not saturate. The condition ii) provides a method to detect the possibility of saturation. The condition iii) implies that if the signal is so small, then it is quantized as zero. Based on the quantizer $\bar{q}^{i}_{j}$ and Assumption \ref{asn-3}, the quantizer applied in this paper is of the form:
\begin{equation}
\label{eqn-6}
q^{i}_{j}(\mu^{i}_{j}, z^{i}_{j})=\mu^{i}_{j}\bar{q}^{i}_{j}(z^{i}_{j}/\mu^{i}_{j}), \quad j\in\{1, \ldots, \ell_{i}\},
\end{equation}
where $\mu^{i}_{j}>0$ is a time-varying quantization parameter.

\begin{assumption}[\cite{Heemels2009networked}]
\label{asn-4}
The initial state $(x_{\p 0}, x_{\rf 0}, x_{\ct 0})$ is known \emph{a priori} and bounded. The quantization parameter $\mu^{i}_{j}$ is such that $|z^{i}_{j}|\leq\mathfrak{m}^{i}_{j}\mu^{i}_{j}$ for all $j\in\{1, \ldots, \ell_{i}\}$ and $i\in\mathcal{N}$.
\end{assumption}

Assumption \ref{asn-4} is to ensure that the quantizer does not saturate. This assumption is enforced easily for linear systems \cite{Nesic2009unified}. See \cite{Franci2010quantised} for more details for the nonlinear case.

For the $i$-th network, combining all the quantizers in $\ell_{i}$ nodes yields the overall quantizer: $q_{i}(\mu_{i}, z_{i}):=(q^{i}_{1}(\mu^{i}_{1}, z^{i}_{1}), \ldots, q^{i}_{\ell_{i}}(\mu^{i}_{\ell_{i}}, z^{i}_{\ell_{i}}))$, where $\mu_{i}:=(\mu^{i}_{1}, \ldots, \mu^{i}_{\ell_{i}})\in\mathbb{R}^{\ell_{i}}$ is evolving as
\begin{align}
\label{eqn-7}
\dot{\mu}_{i}(t)&=0, \quad t\in(r^{i}_{j}, r^{i}_{j+1}), \\
\label{eqn-8}
\mu_{i}({r^{i}_{j}}^{+})&=\Omega_{i}\mu_{i}(r^{i}_{j}), \quad \Omega_{i}:=\diag\{\Omega^{i}_{1}, \ldots, \Omega^{i}_{\ell_{i}}\},
\end{align}
where $\Omega^{i}_{j}\in(0, 1]$. The quantized measurement is defined as $\bar{\mathbf{z}}_{i}=(\bar{y}^{i}_{\p}, \bar{y}^{i}_{\rf}, \bar{y}^{i}_{\ct}, \bar{y}^{i}_{\f}):=(q_{i}(\mu_{i}, y^{i}_{\p}), q_{i}(\mu_{i}, y^{i}_{\rf}), q_{i}(\mu_{i}, y^{i}_{\ct}),$ $ q_{i}(\mu_{i}, y^{i}_{\f}))$; the quantization error is defined as $\epsilon_{i}:=(\epsilon^{i}_{\p}, \epsilon^{i}_{\rf}, \epsilon^{i}_{\ct},$ $\epsilon^{i}_{\f})=(\bar{y}^{i}_{\p}-y^{i}_{\p}, \bar{y}^{i}_{\rf}-y^{i}_{\rf}, \bar{u}^{i}_{\ct}-u^{i}_{\ct}, \bar{u}^{i}_{\f}-u^{i}_{\f})$.

To reduce the transmission frequency, a local ETM is implemented for each network. That is, at each sampling time $t^{i}_{j}$, only when the event-triggered condition for the $i$-th network is satisfied can the quantized measurement be transmitted. Denote by $\hat{z}:=(\hat{y}_{\p}, \hat{y}_{\rf}, \hat{u}_{\ct}, \hat{u}_{\f})\in\mathbb{R}^{n_{z}}$ the received measurement after the transmission, and the control input received by the plant is $\hat{u}:=\hat{u}_{\ct}+\hat{u}_{\f}$. The network-induced errors are defined as $e_{\p}:=\hat{y}_{\p}-y_{\p}$, $e_{\rf}:=\hat{y}_{\rf}-y_{\rf}$, $e_{\ct}:=\hat{u}_{\ct}-u_{\ct}$ and $e_{\f}:=\hat{u}_{\f}-u_{\f}$. From $N$ networks, we denote $\hat{z}=(\hat{z}_{1}, \ldots, \hat{z}_{N})$ and $\mathbf{e}:=\hat{z}-z=(\mathbf{e}_{1}, \ldots, \mathbf{e}_{N})\in\mathbb{R}^{n_{z}}$.

In the interval $[r^{i}_{j}, r^{i}_{j+1}]$, the received measurement $\hat{z}_{i}$ via the $i$-th network is assumed to be implemented via the zero-order hold (ZOH) mechanism, that is,
\begin{align}
\label{eqn-9}
\dot{\hat{z}}_{i}(t)&=0, \quad \forall t\in\left[r^{i}_{j}, r^{i}_{j+1}\right].
\end{align}
At $r^{i}_{j}$, whether $\hat{z}_{i}$ is updated via the latest information is based on the local ETM at $t^{i}_{j}$. Assume that the event-triggered condition for the $i$-th network is given by $\Gamma_{i}\geq0$, where the function $\Gamma_{i}: \mathbb{R}_{\geq0}\rightarrow\mathbb{R}$ will be designed explicitly in Subsection \ref{subsec-DETM}. $\Gamma_{i}\geq0$ implies that the quantized measurement needs to be transmitted, and $\hat{z}_{i}$ is updated with the latest measurement. That is, $\hat{z}_{i}$ is updated by
\begin{align}
\label{eqn-10}
&\hat{z}_{i}({r^{i}_{j}}^{+})=\left\{\begin{aligned}
&\bar{\mathbf{z}}_{i}(r^{i}_{j})+\mathbf{h}^{i}_{z}(\kappa_{i}(t^{i}_{j}), \mathbf{e}_{i}(t^{i}_{j})), &\Gamma_{i}(t^{i}_{j})\geq0, \\
&\hat{z}_{i}(r^{i}_{j}), &\Gamma_{i}(t^{i}_{j})<0,
\end{aligned}\right.
\end{align}
where $\kappa_{i}: \mathbb{R}_{\geq0}\rightarrow\mathbb{N}$ is a counter to record the number of the successful transmission events. That is, $\kappa_{i}({t^{i}_{j}}^{+})=\kappa_{i}(t^{i}_{j})+1$ if $\Gamma_{i}(t^{i}_{j})\geq0$, and $\kappa_{i}({t^{i}_{j}}^{+})=\kappa_{i}(t^{i}_{j})$ otherwise. $\mathbf{h}^{i}_{z}\in\mathbb{R}^{n_{z}}$ is the update function and depends on the time-scheduling protocol as in Subsection \ref{subsec-protocol}. Denote $\mathbf{h}^{i}_{z}:=(\mathbf{h}^{i}_{\p}, \mathbf{h}^{i}_{\rf}, \mathbf{h}^{i}_{\ct}, \mathbf{h}^{i}_{\f})$ from the definition of $\hat{z}_{i}$.
Furthermore, we can rewrite \eqref{eqn-10} as
\begin{align}
\label{eqn-11}
\hat{z}_{i}({r^{i}_{j}}^{+})&=(1-\Upsilon(\Gamma_{i}(t^{i}_{j})))\hat{z}_{i}(r^{i}_{j})+\Upsilon(\Gamma_{i}(t^{i}_{j}))[\bar{\mathbf{z}}_{i}(r^{i}_{j}) \nonumber \\
&\quad \left. +\mathbf{h}^{i}_{z}(\kappa_{i}(t^{i}_{j}), \mathbf{e}_{i}(t^{i}_{j}))\right],
\end{align}
where $\Upsilon: \mathbb{R}\rightarrow\{0, 1\}$ is defined as $\Upsilon(\Gamma_{i})=1$ if $\Gamma_{i}\geq0$ and $\Upsilon(\Gamma_{i})=0$ otherwise. From \eqref{eqn-11}, $\mathbf{e}_{i}$ is updated by
\begin{align*}
&\mathbf{e}_{i}({r^{i}_{j}}^{+})=\hat{z}_{i}({r^{i}_{j}}^{+})-z_{i}({r^{i}_{j}}^{+})\\
&=\mathbf{e}_{i}(r^{i}_{j})+\Upsilon(\Gamma_{i}(t^{i}_{j})) \\
&\quad \times[\mathfrak{h}^{i}_{z}(\kappa_{i}(t^{i}_{j}), \mathbf{x}_{i}(t^{i}_{j}), \mathbf{e}_{i}(t^{i}_{j}), \mu_{i}(t^{i}_{j}))+\bar{\mathbf{z}}_{i}(r^{i}_{j})-\hat{z}_{i}(r^{i}_{j})] \\
&=\mathbf{e}_{i}(r^{i}_{j})+\Upsilon(\Gamma_{i}(t^{i}_{j})) \\
&\quad \times[\mathfrak{h}^{i}_{z}(\kappa_{i}(t^{i}_{j}), \mathbf{x}_{i}(t^{i}_{j}), \mathbf{e}_{i}(t^{i}_{j}), \mu_{i}(t^{i}_{j}))-\mathbf{e}_{i}(t^{i}_{j})],
\end{align*}
where $\mathfrak{h}^{i}_{z}(\kappa_{i}, x_{i}, \mathbf{e}_{i}, \mu_{i})=\epsilon_{i}+\mathbf{h}^{i}_{z}(\kappa_{i}, \mathbf{e}_{i})$ and $\mathbf{x}_{i}=(x^{i}_{\p}, x^{i}_{\rf}, x^{i}_{\ct}, u^{i}_{\f})$.

\subsection{Time-Scheduling Protocols}
\label{subsec-protocol}

Since each network has $\ell_{i}$ nodes with $i\in\mathcal{N}$, the time-scheduling protocol is introduced to decide the node to access to the network. Similar to the analysis and the terminology in \cite{Heemels2010networked}, the function $\mathbf{h}^{i}_{z}(\kappa_{i}, \mathbf{e}_{i})$ in \eqref{eqn-11} is treated as the \textit{protocol}. Based on $\ell_{i}$ nodes for the $i$-th network, $\mathbf{e}_{i}$ is partitioned into $\mathbf{e}_{i}=(\mathbf{e}^{i}_{1}, \ldots, \mathbf{e}^{i}_{\ell_{i}})$. If the $l_{i}$-th node is granted to access to the $i$-th network, where $l_{i}\in\{1, \ldots, \ell_{i}\}$, then the corresponding component $\mathbf{e}^{i}_{l_{i}}$ is updated and the other components are kept constant. In the literature \cite{Heemels2010networked, Nesic2004input}, many time-scheduling protocols can be modeled as $\mathbf{h}^{i}_{z}(\kappa_{i}, \mathbf{e}_{i})$, and two commonly-used protocols are recalled.

The first one is the Round-Robin (RR) protocol \cite{Nesic2004input}. The period of the RR protocol is $\ell_{i}$, and each node has one and only one chance to access to the $i$-th network in a period. The function $\mathbf{h}^{i}_{z}$ is given by
\begin{equation*}
\mathbf{h}^{i}_{z}(\kappa_{i}, \mathbf{e}_{i}):=(I-\Psi_{i}(\kappa_{i}))\mathbf{e}_{i}(t^{i}_{j})+\Psi_{i}(\kappa_{i})\mathbf{\epsilon}_{i}(t^{i}_{j}),
\end{equation*}
where, $\Psi_{i}(\kappa_{i})=\diag\{\Psi^{i}_{1}(\kappa_{i}), \ldots, \Psi^{i}_{\ell_{i}}(\kappa_{i})\}$ and $\Psi^{i}_{l_{i}}(\kappa_{i})\in\mathbb{R}^{n_{l_{i}}\times n_{l_{i}}}$, $\sum^{\ell_{i}}_{l_{i}=1}n_{l_{i}}=n^{i}_{\mathbf{e}}$. $\Psi^{i}_{l_{i}}(\kappa_{i})=I$ if $\kappa_{i}=l_{i}+\jmath\ell_{i}$ with $\jmath\in\mathbb{N}$ and $l_{i}\in\{1, \ldots, \ell_{i}\}$; otherwise, $\Psi^{i}_{l_{i}}(\kappa_{i})=0$.

The second one is Try-Once-Discard (TOD) protocol \cite{Walsh2002stability}. For the TOD protocol, the node with a minimum index where the norm of the local network-induced error is the largest is to access to the network. The function $\mathbf{h}^{i}_{z}$ is given by
\begin{equation*}
\mathbf{h}^{i}_{z}(\kappa_{i}, \mathbf{e}_{i}):=(I-\Psi_{i}(\mathbf{e}_{i}))\mathbf{e}_{i}(t^{i}_{j})+\Psi_{i}(\mathbf{e}_{i})\mathbf{\epsilon}_{i}(t^{i}_{j}),
\end{equation*}
where, $\Psi_{i}(\mathbf{e}_{i})=\diag\{\Psi^{i}_{1}(\mathbf{e}_{i}), \ldots, \Psi^{i}_{\ell_{i}}(\mathbf{e}_{i})\}$, and $\Psi^{i}_{l_{i}}(\mathbf{e}_{i})=I$ if $\min\left\{\arg\max_{1\leq k\leq \ell_{i}}|\mathbf{e}^{i}_{k}|\right\}=l_{i}$; otherwise, $\Psi^{i}_{l_{i}}(\mathbf{e}_{i})=0$.

\section{Development of Hybrid Model}
\label{sec-developandreformulate}

After the presentation of the information transmission, we construct the hybrid model for the event-triggered tracking control of NQCS in this section. Since our objective is to guarantee the convergence of $x_{\p}$ towards $x_{\rf}$, we define the tracking error $\eta:=x_{\p}-x_{\rf}\in\mathbb{R}^{n_{\p}}$, and the auxiliary variable $e_{\ax}:=(e_{\eta}, e_{\ct}):=(e_{\p}-e_{\rf}, e_{\ct})\in\mathbb{R}^{n_{\ax}}$ with the network-induced errors $e_{\p}, e_{\rf}, e_{\ct}$ defined in Section \ref{sec-problemformation}, where $n_{\ax}=n_{y}+n_{\ct}$. Combining all the variables and analyses in Subsection \ref{subsec-informationtransmission}, we derive the following impulsive model:
\begin{subequations}
\label{eqn-12}
\begin{align}
\label{eqn-12a}
&\left.\begin{aligned}
\dot{\eta}&=F_{\eta}(\delta, \eta, x_{\ct}, x_{\rf}, e_{\ax}, e_{\f}, e_{\rf})\\
\dot{x}_{\ct}&=F_{\ct}(\delta, \eta, x_{\ct}, x_{\rf}, e_{\ax}, e_{\f}, e_{\rf})\\
\dot{x}_{\rf}&=F_{\rf}(\delta, x_{\rf}, e_{\f}), \quad \dot{\mu}=0\\
\dot{e}_{\ax}&=G_{\ax}(\delta, \eta, x_{\ct}, x_{\rf}, e_{\ax}, e_{\f}, e_{\rf})\\
\dot{e}_{\rf}&=G_{\rf}(\delta, \eta, x_{\ct}, x_{\rf}, e_{\ax}, e_{\f}, e_{\rf})\\
\dot{e}_{\f}&=G_{\f}(\delta, \eta, x_{\ct}, x_{\rf}, e_{\ax}, e_{\f}, e_{\rf}) \\
\end{aligned}\right\} \quad t^{i}\in[r^{i}_{j}, r^{i}_{j+1}], \\
\label{eqn-12b}
&\begin{aligned}
\mu_{i}({r^{i}_{j}}^{+})&=\Omega_{i}\mu_{i}(r^{i}_{j}), \quad \delta_{i}({r^{i}_{j}}^{+})=\delta_{i}(r^{i}_{j}), \\
e^{i}_{\ax}({r^{i}_{j}}^{+})&=e^{i}_{\ax}(r^{i}_{j})+\Upsilon(\Gamma_{i}(t^{i}_{j}))[-e^{i}_{\ax}(t^{i}_{j})  \\
&\quad h^{i}_{\ax}(\kappa_{i}(t^{i}_{j}), x_{i}(t^{i}_{j}), e_{i}(t^{i}_{j}), \mu_{i}(t^{i}_{j}))],\\
e^{i}_{\rf}({r^{i}_{j}}^{+})&=e^{i}_{\rf}(r^{i}_{j})+\Upsilon(\Gamma_{i}(t^{i}_{j}))[-e^{i}_{\rf}(t^{i}_{j})  \\
&\quad +h^{i}_{\rf}(\kappa_{i}(t^{i}_{j}), x_{i}(t^{i}_{j}), e_{i}(t^{i}_{j}), \mu_{i}(t^{i}_{j}))],\\
e^{i}_{\f}({r^{i}_{j}}^{+})&=e^{i}_{\f}(r^{i}_{j})+\Upsilon(\Gamma_{i}(t^{i}_{j}))[-e^{i}_{\f}(t^{i}_{j})  \\
&\quad +h^{i}_{\f}(\kappa_{i}(t^{i}_{j}), x_{i}(t^{i}_{j}), e_{i}(t^{i}_{j}), \mu_{i}(t^{i}_{j}))],
\end{aligned}
\end{align}
\end{subequations}
where $e_{i}:=(e^{i}_{\ax}, e^{i}_{\rf}, e^{i}_{\f})$, $h^{i}_{\ax}=(\mathfrak{h}^{i}_{\p}-\mathfrak{h}^{i}_{\rf}, \mathfrak{h}^{i}_{\ct}), h^{i}_{\rf}=\mathfrak{h}^{i}_{\rf}$ and $h^{i}_{\f}=\mathfrak{h}^{i}_{\f}$. $\delta:=(\delta_{1}, \ldots, \delta_{N})\in\mathbb{R}^{N}$, and $\delta_{i}\in\mathbb{R}_{\geq0}$ is to model the `continuous' time of the $i$-th network and depends on $u^{i}_{\f}$ and/or its differential \cite{Postoyan2014tracking, Ren2019tracking}. All the functions in \eqref{eqn-12a} can be derived by relevant calculations. Now, our objective is to derive reasonable  conditions and ETMs simultaneously to guarantee ISS of the system \eqref{eqn-12} from $(e_{\rf}, e_{\f})$ to $(\eta, e_{\ax})$.

To facilitate the analysis afterwards, we transform \eqref{eqn-12} into a hybrid model in the formalism of \cite{Cai2009characterizations}. Define $x:=(\eta, x_{\ct}, x_{\rf})\in\mathbb{R}^{n_{x}}$ and $e:=(e_{\ax}, e_{\rf}, e_{\f})\in\mathbb{R}^{n_{e}}$ with $n_{x}=n_{\p}+n_{\ct}+n_{\rf}$ and $n_{e}=n_{\ax}+n_{y}+n_{u}$. Define $m:=(m_{1}, \ldots, m_{N})\in\mathbb{R}^{n_{e}}$ with $m_{i}:=h_{i}(\kappa_{i}, e_{i})-e_{i}\in\mathbb{R}^{n^{i}_{e}}$ storing the information for the update, where $e_{i}:=(e^{i}_{\ax}, e^{i}_{\rf}, e^{i}_{\f})$ and $h_{i}:=(h^{i}_{\ax}, h^{i}_{\rf}, h^{i}_{\f})$ are defined in \eqref{eqn-12}. Let $\kappa:=(\kappa_{1}, \ldots, \kappa_{N})\in\mathbb{R}^{N}$ with $\kappa_{i}\in\mathbb{N}$ defined in \eqref{eqn-10}; $\mu:=(\mu_{1}, \ldots, \mu_{N})\in\mathbb{R}^{\mathfrak{L}}$ with $\mathfrak{L}:=\sum_{i\in\mathcal{N}}\ell_{i}$ and $\mu_{i}\in\mathbb{R}$ defined in \eqref{eqn-7}-\eqref{eqn-8}; $\tau:=(\tau_{1}, \ldots, \tau_{N})\in\mathbb{R}^{N}$ with $\tau_{i}\in\mathbb{R}$ defined as a timer to record both sampling intervals and transmission delays for the $i$-th network; $b:=(b_{1}, \ldots, b_{N})\in\mathbb{R}^{N}$ with $b_{i}\in\{0, 1\}$ to show whether the next jump is from the sampling event or the update event. That is, for the $i$-th network, $b_{i}=0$ means that the next event will be the sampling event, while $b_{i}=1$ means that the next event will be the update event. Denote $\mathfrak{X}:=(x, e, \mu, m, \delta, \tau, \kappa, b)\in\mathscr{R}:=\mathbb{R}^{n_{x}}\times\mathbb{R}^{n_{e}}
\times\mathbb{R}^{\mathfrak{L}}\times\mathbb{R}^{n_{e}}\times\mathbb{R}^{N}\times\mathbb{R}^{N}\times\mathbb{R}^{N}\times\{0, 1\}^{N}$, and the hybrid model is given by
\begin{align}
\label{eqn-13}
\left\{\begin{aligned}
&\dot{\mathfrak{X}}=F(\mathfrak{X}), &\quad& \mathfrak{X}\in C,  \\
&\mathfrak{X}^{+}=G(\mathfrak{X}), &\quad& \mathfrak{X}\in D,
\end{aligned}\right.
\end{align}
where $C:=\bigcup^{N}_{i=1} C_{i}$, $D:=\bigcup^{N}_{i=1}(D_{1i}\cup D_{2i})$,
\begin{align}
C_{i}&:=\{\mathfrak{X}\in\mathscr{R}: (b_{i}, \tau_{i})\in(\{0\}\times\mathbf{T}_{i})\cup(\{1\}\times\mathbf{\Delta}_{i})\}, \nonumber \\
D_{1i}&:=\{\mathfrak{X}\in\mathscr{R}: (b_{i}, \tau_{i})\in\{0\}\times[\varepsilon_{i}, T_{i}]\}, \nonumber \\
D_{2i}&:=\{\mathfrak{X}\in\mathscr{R}: (b_{i}, \tau_{i})\in\{1\}\times\mathbf{\Delta}_{i}\}, \nonumber
\end{align}
and $\mathbf{T}_{i}:=[0, T_{i}], \mathbf{\Delta}_{i}:=[0, \Delta_{i}]$ with $T_{i}>0$ and $\Delta_{i}\geq0$ from Assumption \ref{asn-2}. The mapping $F$ in \eqref{eqn-13} is defined as
\begin{align}
\label{eqn-14}
F(\mathfrak{X})&:=(f(\delta, x, e), g(\delta, x, e), 0, 0, \mathds{E}, \mathds{E}, 0, 0),
\end{align}
where $f:=(F_{\eta}, F_{\rf}, F_{\ct})$ and $g:=(G_{\ax}, G_{\rf}, G_{\f})$ are derived from \eqref{eqn-12a}. The mapping $G$ in \eqref{eqn-13} is defined as
\begin{align}
\label{eqn-15}
G(\mathfrak{X})&:=\left\{\begin{aligned}
&G_{1}(\mathfrak{X}), &\quad& \mathfrak{X}\in D_{1}, \\
&G_{2}(\mathfrak{X}), &\quad& \mathfrak{X}\in D_{2},
\end{aligned}\right.
\end{align}
with $G_{1}(\mathfrak{X})=\bigcup^{N}_{i=1}G_{1i}(\mathfrak{X}), D_{1}=\bigcup^{N}_{i=1}D_{1i},  G_{2}(\mathfrak{X})=\bigcup^{N}_{i=1}G_{2i}(\mathfrak{X}), D_{2}=\bigcup^{N}_{i=1}D_{2i}$, and
\begin{align}
\label{eqn-16}
G_{1i}(\mathfrak{X})&:=\left\{\begin{aligned}
&\begin{bmatrix}
x \\ e \\ \mu \\ \mathcal{M}_{1i}(x, e, m, \kappa) \\ \delta \\ \Lambda_{i}\tau \\ \kappa+\Upsilon(\Gamma_{i})(I-\Lambda_{i})\mathds{E} \\ b+(I-\Lambda_{i})\mathds{E}
\end{bmatrix}, &\quad& \mathfrak{X}\in D_{1i}, \\
&\varnothing, &\quad& \mathfrak{X}\notin D_{1i},
\end{aligned}\right.\\
\label{eqn-17}
G_{2i}(\mathfrak{X})&:=\left\{\begin{aligned}
&\begin{bmatrix}
x \\ \mathcal{E}_{i}(x, e, m, \kappa) \\ \boldsymbol{\Omega}_{i}\mu \\ \mathcal{M}_{2i}(x, e, m, \kappa) \\ \delta\\  \tau  \\ \kappa  \\  \Lambda_{i}b
\end{bmatrix}, &\quad& \mathfrak{X}\in D_{2i},  \\
&\varnothing, &\quad& \mathfrak{X}\notin D_{2i},
\end{aligned}\right.
\end{align}
where $\Lambda_{i}:=\diag\{\Lambda^{1}_{i}, \ldots, \Lambda^{N}_{i}\}\in\mathbb{R}^{N\times N}$ with $\Lambda^{k}_{i}=0$ if $k=i\in\mathcal{N}$ and $\Lambda^{k}_{i}=1$ otherwise; $\boldsymbol{\Omega}_{i}:=\diag\{\boldsymbol{\Omega}^{1}_{i}, \ldots, \boldsymbol{\Omega}^{N}_{i}\}\in\mathbb{R}^{\mathfrak{L}\times\mathfrak{L}}$ with $\boldsymbol{\Omega}^{k}_{i}=\Omega_{i}$ if $k=i\in\mathcal{N}$ and $\boldsymbol{\Omega}^{k}_{i}=I$ otherwise;
\begin{align*}
\mathcal{M}_{1i}(x, e, m, \kappa)&:=\Phi_{i}m+(I-\Phi_{i})M_{1i}(x, e, m, \kappa), \\
\mathcal{M}_{2i}(x, e, m, \kappa)&:=\Phi_{i}m+(I-\Phi_{i})M_{2i}(e, m), \\
\mathcal{E}_{i}(x, e, m, \kappa)&:=\Phi_{i}m+\Upsilon(\Gamma_{i})(I-\Phi_{i})E_{i}(e, m).
\end{align*}
Here, $\Phi_{i}:=\diag\{\Phi^{1}_{i}, \ldots, \Phi^{N}_{i}\}\in\mathbb{R}^{n_{e}\times n_{e}}$, $M_{1i}:=(M^{1}_{1i}, \ldots,$ $M^{N}_{1i})\in\mathbb{R}^{n_{e}}$, $M_{2i}:=(M^{1}_{2i}, \ldots, M^{N}_{2i})\in\mathbb{R}^{n_{e}}$ and $E_{i}:=(E^{1}_{i}, \ldots, E^{N}_{i})\in\mathbb{R}^{n_{e}}$. If $k=i$, then $\Phi^{k}_{i}=0$, $M^{k}_{1i}=(1-\Upsilon(\Gamma_{i}))m_{i}+\Upsilon(\Gamma_{i})(h_{i}(\kappa_{i}, e_{i})-e_{i})$, $M^{k}_{2i}=-e_{i}-m_{i}$ and $E^{k}_{i}=e_{i}+m_{i}$. If $k\neq i$, then $\Phi^{k}_{i}=I$ and $M^{k}_{1i}=M^{k}_{2i}=E^{k}_{i}=0$.

For the hybrid model \eqref{eqn-13}, the sets $C$ and $D$ are closed. Since $f_{\p}, f_{\ct}, g_{\p}$ and $g_{\ct}$ are continuous in Subsection \ref{subsec-trackingproblem}, the functions $f$ and $g$ in \eqref{eqn-14} are continuous, and thus the flow map $F$ in \eqref{eqn-14} is continuous. The jump map $G$ in \eqref{eqn-15} is continuous and locally bounded from the continuity of $G_{1i}$ in \eqref{eqn-16} and $G_{2i}$ in \eqref{eqn-17}. Hence, we can verify that the hybrid model \eqref{eqn-13} satisfies the basic assumptions introduced in Section \ref{sec-preliminaries}.

\section{Main Results}
\label{sec-mainresults}

In this section, the main results are established. We first present some necessary assumptions, then design the event-triggered condition for each network, and finally establish the convergence of the tracking error.

\subsection{Assumptions}
\label{subsec-assump}

To begin with, two assumptions are presented for the $(e_{i}, \mu_{i})$-subsystem, and an assumption is given for $x$-subsystem under the designed controller.

\begin{assumption}
\label{asn-5}
There exist a function $W_{i}: \mathbb{R}^{n^{i}_{e}}\times\mathbb{R}^{\ell_{i}}\times\mathbb{R}^{n^{i}_{e}}\times\mathbb{N}\times\{0, 1\}\rightarrow\mathbb{R}_{\geq0}$ which is locally Lipschitz in $(e_{i}, \mu_{i}, m_{i})$ for all $\kappa_{i}\in\mathbb{N}, b_{i}\in\{0, 1\}$, $\alpha_{\jmath i}\in\mathcal{K}_{\infty}$, $\jmath\in\{1, \ldots, 6\}$, and $\lambda_{i}\in[0, 1)$ such that for all $(e_{i}, \mu_{i}, m_{i}, \kappa_{i}, b_{i})\in\mathbb{R}^{n^{i}_{e}}\times\mathbb{R}^{\ell_{i}}\times\mathbb{R}^{n^{i}_{e}}\times\mathbb{N}\times\{0, 1\}$,
\begin{align}
\label{eqn-18}
&\alpha_{1i}(|e^{i}_{\ax}|)\leq W_{i}(e_{i}, \mu_{i}, m_{i}, \kappa_{i}, b_{i})\leq\alpha_{2i}(|e_{i}|),\\
\label{eqn-19}
&W_{i}(e_{i}, \mu_{i}, h_{i}(\kappa_{i}, e_{i})-e_{i}, \kappa_{i}+1, 1) \nonumber \\
&\quad \leq\lambda_{i}W_{i}(e_{i}, \mu_{i}, m_{i}, \kappa_{i}, 0)+\alpha_{3i}(|e^{i}_{\f}|)+\alpha_{4i}(|e^{i}_{\rf}|), \\
\label{eqn-20}
&W_{i}(e_{i}+m_{i}, \Omega_{i}\mu_{i}, -e_{i}-m_{i}, \kappa_{i}, 0)\nonumber \\
&\quad \leq W_{i}(e_{i}, \mu_{i}, m_{i}, \kappa_{i}, 1)+\alpha_{5i}(|e^{i}_{\f}|)+\alpha_{6i}(|e^{i}_{\rf}|).
\end{align}
\end{assumption}

\begin{assumption}
\label{asn-6}
There exist a continuous function $H_{ib_{i}}: \mathbb{R}^{n_{x}}\times\mathbb{R}^{n_{e}}\rightarrow\mathbb{R}_{>0}$, $\sigma_{1ib_{i}}, \sigma_{2ib_{i}}\in\mathcal{K}_{\infty}$ and $L_{ib_{i}}\in[0, \infty)$ such that for all $(x, \kappa_{i}, b_{i})\in\mathbb{R}^{n_{x}}\times\mathbb{N}
\times\{0, 1\}$ and almost all $(e_{i}, \mu_{i}, m_{i})\in\mathbb{R}^{n^{i}_{e}}\times\mathbb{R}^{\ell_{i}}\times\mathbb{R}^{n^{i}_{e}}$,
\begin{align}
\label{eqn-21}
&\left\langle\frac{\partial W_{i}(e_{i}, \mu_{i}, m_{i}, \kappa_{i}, b_{i})}{\partial e_{i}}, g_{i}(\delta, x, e)\right\rangle\leq H_{ib_{i}}(x, e)   \nonumber \\
& +L_{ib_{i}}W_{i}(e_{i}, \mu_{i}, m_{i}, \kappa_{i}, b_{i})+\sigma_{1ib_{i}}(|e^{i}_{\f}|)+\sigma_{2ib_{i}}(|e^{i}_{\rf}|).
\end{align}
\end{assumption}

Assumptions \ref{asn-5}-\ref{asn-6} are on the $e_{i}$-subsystem, whose properties are described via the function $W_{i}$. Assumption \ref{asn-5} is to estimate the jumps of $W_{i}$ at the discrete-time instants. Specifically, \eqref{eqn-19} is for the successful transmission jumps (i.e., $\Gamma_{i}\geq0$) at the sampling instants $t^{i}_{j}$, and \eqref{eqn-20} is for the update jumps at the arrival instants $r^{i}_{j}$. Assumption \ref{asn-6} is to estimate the derivative of $W_{i}$ in the continuous-time intervals, and the coupling is shown via the function $H_{ib_{i}}$. Since Assumptions \ref{asn-5}-\ref{asn-6} are applied to the $e_{i}$-subsystem, \eqref{eqn-19}-\eqref{eqn-20} hold with respect to the additional items $e^{i}_{\rf}$ and $e^{i}_{\f}$, which are parts of $e_{i}$ and treated as the internal disturbances caused by the network. Similar conditions have been considered in existing works \cite{Heemels2010networked, Ren2019tracking, Postoyan2014tracking}, where however only a common communication network and TTC are studied.

\begin{assumption}
\label{asn-7}
There exist a locally Lipschitz function $V: \mathbb{R}^{n_{x}}\rightarrow\mathbb{R}_{\geq0}$, $\alpha_{1V}, \alpha_{2V}, \zeta_{1ib_{i}}, \zeta_{2ib_{i}}, \zeta_{3ib_{i}}, \zeta_{4ib_{i}}\in\mathcal{K}_{\infty}$, and $\mu, \theta_{ib_{i}}, \gamma_{ib_{i}}>0, \bar{L}_{ib_{i}}\in\mathbb{R}$ such that
\begin{align}
\label{eqn-22}
&\alpha_{1V}(|\eta|)\leq V(x)\leq\alpha_{2V}(|x|), \quad \forall x\in\mathbb{R}^{n_{x}},
\end{align}
and for all $(e_{i}, \mu_{i}, m_{i}, \kappa_{i}, b_{i})\in\mathbb{R}^{n^{i}_{e}}\times\mathbb{R}^{\ell_{i}}\times\mathbb{R}^{n^{i}_{e}}\times\mathbb{N}\times\{0, 1\}$ and almost all $x\in\mathbb{R}^{n_{x}}$,
\begin{align}
\label{eqn-23}
&\langle\nabla V(x), f(\delta, x, e)\rangle\leq-\mu V(x)-\sum^{N}_{i=1}\left[H^{2}_{ib_{i}}(x, e)\right.  \nonumber\\
&\quad +(\gamma^{2}_{ib_{i}}-\theta_{ib_{i}})W^{2}_{i}(e_{i}, \mu_{i}, m_{i}, \kappa_{i}, b_{i})-K_{ib_{i}}(x, e, \mu, m)  \nonumber\\
&\quad \left.-\varphi_{ib_{i}}(z_{i})+\zeta_{1ib_{i}}(|e^{i}_{\f}|)+\zeta_{2ib_{i}}(|e^{i}_{\rf}|)\right], \\
\label{eqn-24}
&\langle\nabla\varphi_{ib_{i}}(z_{i}), f^{i}_{z}(\delta, x, e)\rangle\leq\bar{L}_{ib_{i}}\varphi_{i}(z_{i})+K_{ib_{i}}(x, e, \mu, m)  \nonumber \\
&\quad +H^{2}_{ib_{i}}(x, e)+\zeta_{3ib_{i}}(|e^{i}_{\f}|)+\zeta_{4ib_{i}}(|e^{i}_{\rf}|),
\end{align}
where $H_{ib_{i}}$ is defined in Assumption \ref{asn-6}, $\varphi_{ib_{i}}: \mathbb{R}^{n^{i}_{z}}\rightarrow\mathbb{R}_{\geq0}$ is a locally Lipschitz function with $\varphi_{ib_{i}}(0)=0$, and $K_{ib_{i}}: \mathbb{R}^{n_{x}}\times\mathbb{R}^{n_{e}}\times\mathbb{R}^{\mathfrak{L}}\times\mathbb{R}^{n_{e}}\rightarrow\mathbb{R}_{\geq0}$ is a continuous function.
\end{assumption}

Assumption \ref{asn-7} describes the property of the $x$-subsystem via the function $V$. Under the designed controller, \eqref{eqn-22}-\eqref{eqn-23} imply that the $\eta$-subsystem satisfies the ISS-like property from $(\sum^{N}_{i=1}W_{i}, e_{\f}, e_{\rf})$ to $\eta$. This assumption is reasonable due to the implementation of the emulation-based approach, where the controller is assumed to be known \textit{a priori} to ensure the tracking performance robustly in the network-free case. Hence, in the presence of the network, $(\sum^{N}_{i=1}W_{i}, e_{\f}, e_{\rf})$ is treated as a whole disturbance from the interior of the plant. Moreover, \eqref{eqn-24} provides the growth bound on the derivative of the function $\varphi_{ib_{i}}$ on the flow. Note that the information of multiple networks is not required in Assumption \ref{asn-7}, and that the function $\varphi_{ib_{i}}$ will be applied to design the ETMs. Finally, for the linear case, Assumptions \ref{asn-5}-\ref{asn-7} can be represented as a whole linear matrix inequality; see e.g. \cite{Wang2019periodic}.

\subsection{Decentralized Event-Triggered Mechanisms}
\label{subsec-DETM}

With Assumption \ref{asn-5}-\ref{asn-7}, we next show how to design the ETM for each network based on the functions $W_{i}$ and $V$. To this end, the function $\Gamma_{i}$ in \eqref{eqn-11} is defined as a mapping from $\mathbb{R}^{n^{i}_{z}}\times\mathbb{R}^{n^{i}_{e}}\times
\mathbb{R}^{\ell_{i}}\times\mathbb{R}^{n^{i}_{e}}\times\mathbb{N}\times\{0, 1\}$ to $\mathbb{R}$:
\begin{align}
\label{eqn-25}
\Gamma_{i}(z_{i}, e_{i}, \mu_{i}, m_{i}, \kappa_{i}, b_{i})&:=(1-2b_{i})\gamma_{ib_{i}}W^{2}_{i}(e_{i}, \mu_{i}, m_{i}, \kappa_{i}, b_{i}) \nonumber \\
&\quad -(1-b_{i})\rho_{i}\bar{\lambda}_{i}\varphi_{ib_{i}}(z_{i}),
\end{align}
where $W_{i}$ is defined in Assumption \ref{asn-5}, $\varphi_{ib_{i}}$ is defined in Assumption \ref{asn-7}, $\rho_{i}\in\mathbb{R}_{\geq0}$ is a design parameter satisfying $\rho_{i}\in[0, \bar{\rho}_{i})$, and
\begin{align}
\label{eqn-26}
\bar{\lambda}_{i}&:=\max\left\{\lambda_{i}, \frac{\rho_{i}\gamma_{i0}}{1-\rho_{i}\bar{L}_{i0}}\right\}, \\
\label{eqn-27}
\bar{\rho}_{i}&:=\left\{\begin{aligned}
&1, &\quad& \bar{L}_{i0}\leq-\gamma_{i0},  \\
&\min\left\{1, (\bar{L}_{i0}+\gamma_{i0})^{-1}\right\}, &\quad& \bar{L}_{i0}>-\gamma_{i0},
\end{aligned}\right.
\end{align}
with $\lambda_{i}$ in Assumption \ref{asn-5} and $\gamma_{i0}, \bar{L}_{i0}$ in Assumption \ref{asn-7}.

With the function \eqref{eqn-25}, the event-triggered condition is $\Gamma_{i}(z_{i}, e_{i}, \mu_{i}, m_{i}, \kappa_{i}, b_{i})\geq0$. The proposed event-triggered condition is similar to those in \cite{Wang2019periodic, Tabuada2007event, Dimarogonas2011distributed} for the ETC in different contexts. One difference between \eqref{eqn-25} and the existing ones lies in the local logical variable $b_{i}$, which leads to two cases in \eqref{eqn-25}. Since the case $b_{i}=1$ implies that the update event will occur at the arrival instant, the ETM is not needed and $\Gamma_{i}(z_{i}, e_{i}, \mu_{i}, m_{i}, \kappa_{i}, 1)=-\gamma_{i1}W^{2}_{i}(e_{i}, \mu_{i}, m_{i}, \kappa_{i}, 1)<0$, which thus implies that the ETM will not be implemented in this case. In contrast, for the case $b_{i}=0$, the next event is the transmission event, and the ETM is implemented to determine whether the sampled measurement will be transmitted. Hence, $\Gamma_{i}(z_{i}, e_{i}, \mu_{i}, m_{i}, \kappa_{i}, 0)=\gamma_{i0}W^{2}_{i}(e_{i}, \mu_{i}, m_{i}, \kappa_{i}, 0)-\rho_{i}\bar{\lambda}_{i}\varphi_{i0}(z_{i})\geq0$ will be verified in this case. As a result, the parameters in \eqref{eqn-26}-\eqref{eqn-27} only depend on the case $b_{i}=0$, and all the designed event-triggered conditions are consistent with the transmission setup and decentralized since only local information is involved in each event-triggered condition.

\begin{remark}
In \eqref{eqn-25}, if $\rho_{i}\equiv0$ for some $i\in\mathcal{N}$, then $\Gamma_{i}$ is always positive, and thus the proposed ETC is reduced to the TTC as in \cite{Ren2019tracking}, where $T_{i}$ is called the maximally allowable transmission interval. Since all the networks are independent, both TTC and ETC can be combined by allowing that some networks perform the TTC while the others perform the ETC, which is a potential extension of this paper.
\hfill$\square$
\end{remark}

Finally, consider the following differential equation
\begin{align}
\label{eqn-28}
\dot{\phi}_{ib_{i}}&=-2L_{ib_{i}}\phi_{ib_{i}}-\gamma_{ib_{i}}((1+\varrho_{ib_{i}})\phi^{2}_{ib_{i}}+1),
\end{align}
where $i\in\mathcal{N}$, $L_{ib_{i}}\geq0$ is given in Assumption \ref{asn-6}, and $\gamma_{ib_{i}}>0$ is given in Assumption \ref{asn-7}. In \eqref{eqn-28}, $\varrho_{ib_{i}}\in(0, \bar{\lambda}^{-2}_{i}\phi^{-2}_{ib_{i}}(0)-1)$, and thus the initial values $\phi_{ib_{i}}(0)\in(1, \bar{\lambda}^{-1}_{i})$, where $\bar{\lambda}_{i}$ is given in \eqref{eqn-26}. From Claim 1 in \cite{Carnevale2007lyapunov} and Claim 1 in \cite{Postoyan2014tracking}, the solutions to \eqref{eqn-28} are strictly decreasing as long as $\phi_{ib_{i}}\geq0$.

\subsection{Tracking Performance Analysis}
\label{subsec-trackanalyz}

Now we are ready to state the main result of this section.

\begin{theorem}
\label{thm-1}
Consider the system \eqref{eqn-13} and let Assumptions \ref{asn-1}-\ref{asn-7} hold. If the MASP $T_{i}$ and the MAD $\Delta_{i}$ satisfy
\begin{subequations}
\label{eqn-29}
\begin{align}
\label{eqn-29-1}
\gamma_{i0}\phi_{i0}(\tau_{i})&\geq(1+\varrho_{i1})\bar{\lambda}^{2}_{i}\gamma_{i1}\phi_{i1}(0), && \tau_{i}\in\mathbf{T}_{i},\\
\label{eqn-29-2}
\gamma_{i1}\phi_{i1}(\tau_{i})&\geq(1+\varrho_{i0})\gamma_{i0}\phi_{i0}(\tau_{i}), && \tau_{i}\in\mathbf{\Delta}_{i},
\end{align}
\end{subequations}
where $\phi_{ib_{i}}$ is the solution to \eqref{eqn-28} with $\phi_{ib_{i}}(0), \phi_{ib_{i}}(T_{i})>0$, then the system \eqref{eqn-13} is ISS from $(e_{\rf}, e_{\f})$ to $(\eta, e_{\ax})$. That is, there exist $\beta\in\mathcal{KLL}$ and $\varphi_{1}\in\mathcal{K}_{\infty}$ such that for all $(t, j)\in\mathbb{R}_{\geq0}\times\mathbb{N}$,
\begin{align}
\label{eqn-30}
|(\eta(t, j), e_{\ax}(t, j))|&\leq\beta(|\mathfrak{X}(0, 0)|, t, j)+\varphi_{1}(\|e_{\f}\|_{(t, j)}) \nonumber\\
&\quad +\varphi_{2}(\|e_{\rf}\|_{(t, j)}).
\end{align}
\end{theorem}

The proof is omitted due to the space limitation; see \cite{Postoyan2014tracking, Ren2019tracking} for the similar proof strategy. The proof strategy is to construct the Lyapunov function $U(\mathfrak{X}):=V(x)+\sum^{N}_{i=1}\max\{\gamma_{ib_{i}}\phi_{ib_{i}}(\tau_{i})W^{2}_{i}(e_{i}, \mu_{i}, m_{i}, \kappa_{i}, b_{i}), (1-b_{i})\rho_{i}\varphi_{ib_{i}}(z_{i})\}$ based on Assumptions \ref{asn-5}-\ref{asn-7}, then to show that the function $U(\mathfrak{X})$ is decreasing on the flow and non-increasing at the jumps, and finally to guarantee the convergence of $U(\mathfrak{X})$ via hybrid systems theory \cite{Cai2009characterizations}. Theorem \ref{thm-1} implies the convergence of the tracking error to a region around the origin, and the size of the convergence region depends on the network-induced error $(e_{\rf}, e_{\f})$. If the feedforward control inputs are transmitted directly to the plant and reference system, then $e_{\f}=0, \varphi_{1}\equiv0$, and thus the convergence region can be further smaller.

Comparing with previous works \cite{Postoyan2014tracking, Ren2019tracking, Tallapragada2013event, Dolk2016output, Wang2019periodic} on NCS and \cite{Hong2006tracking, Cheng2016event} on MAS, the event-triggered tracking control problem is studied here for NQCS under decentralized ETMs and network constraints. In particular, quantization effects and/or time delays are not considered in \cite{Dolk2016output, Hong2006tracking, Cheng2016event, Tallapragada2013event, Postoyan2014tracking, Wang2019periodic}, and the time-triggered tracking control is addressed in \cite{Postoyan2014tracking, Ren2019tracking}. Therefore, a unified model is developed here and the tracking performance is achieved via less communication, which is shown via the numerical example in the next section.

\section{Numerical Example}
\label{sec-illustration}

Consider two connected single-link robot arms, whose dynamics are presented as ($i=1, 2$)
\begin{align}
\label{eqn-31}
\begin{aligned}
\dot{\mathfrak{q}}^{i1}_{\p}&=\mathfrak{q}^{i2}_{\p}, \\
\dot{\mathfrak{q}}^{i2}_{\p}&=-a_{i}\sin\mathfrak{q}^{i1}_{\p}+\sum^{2}_{j=1}b_{ij}(\mathfrak{q}^{1j}_{\p}-\mathfrak{q}^{2j}_{\p})+c_{i}u_{i},
\end{aligned}
\end{align}
where $\mathfrak{q}^{i}_{\p}:=(\mathfrak{q}^{i1}_{\p}, \mathfrak{q}^{i2}_{\p})\in\mathbb{R}^{2}$ with the configuration coordinate $\mathfrak{q}^{i1}_{\p}$ and the velocity $\mathfrak{q}^{i2}_{\p}$, both of which are measurable, $u_{i}\in\mathbb{R}$ is the input torque, and $a_{i}, c_{i}>0, b_{ij}\in\mathbb{R}$ are certain constants. The references are given by
\begin{align}
\label{eqn-32}
\begin{aligned}
\dot{\mathfrak{q}}^{i1}_{\rf}&=\mathfrak{q}^{i2}_{\rf}, \\
\dot{\mathfrak{q}}^{i2}_{\rf}&=-a_{i}\sin\mathfrak{q}^{i1}_{\rf}+\sum^{2}_{j=1}b_{ij}(\mathfrak{q}^{1j}_{\rf}-\mathfrak{q}^{2j}_{\rf})+c_{i}u^{i}_{\f},
\end{aligned}
\end{align}
where $\mathfrak{q}^{i}_{\rf}:=(\mathfrak{q}^{i1}_{\rf}, \mathfrak{q}^{i2}_{\rf})\in\mathbb{R}^{2}$ are the measurable reference state, and $u^{i}_{\f}=5\sin(5t)$ is the feedforward input. In the network-free case, the feedback controller is designed as $u^{i}_{\ct}=-c^{-1}_{i}[a_{i}(\sin(\mathfrak{q}^{i1}_{\p})-\sin(\mathfrak{q}^{i1}_{\rf}))-(\mathfrak{q}^{i1}_{\p}
-\mathfrak{q}^{i1}_{\rf})-(\mathfrak{q}^{i2}_{\p}-\mathfrak{q}^{i2}_{\rf})]$ such that the tracking error is asymptotically stable.

\begin{figure}[!t]
\begin{center}
\begin{picture}(70,100)
\put(-70,-20){\resizebox{70mm}{40mm}{\includegraphics[width=2.5in]{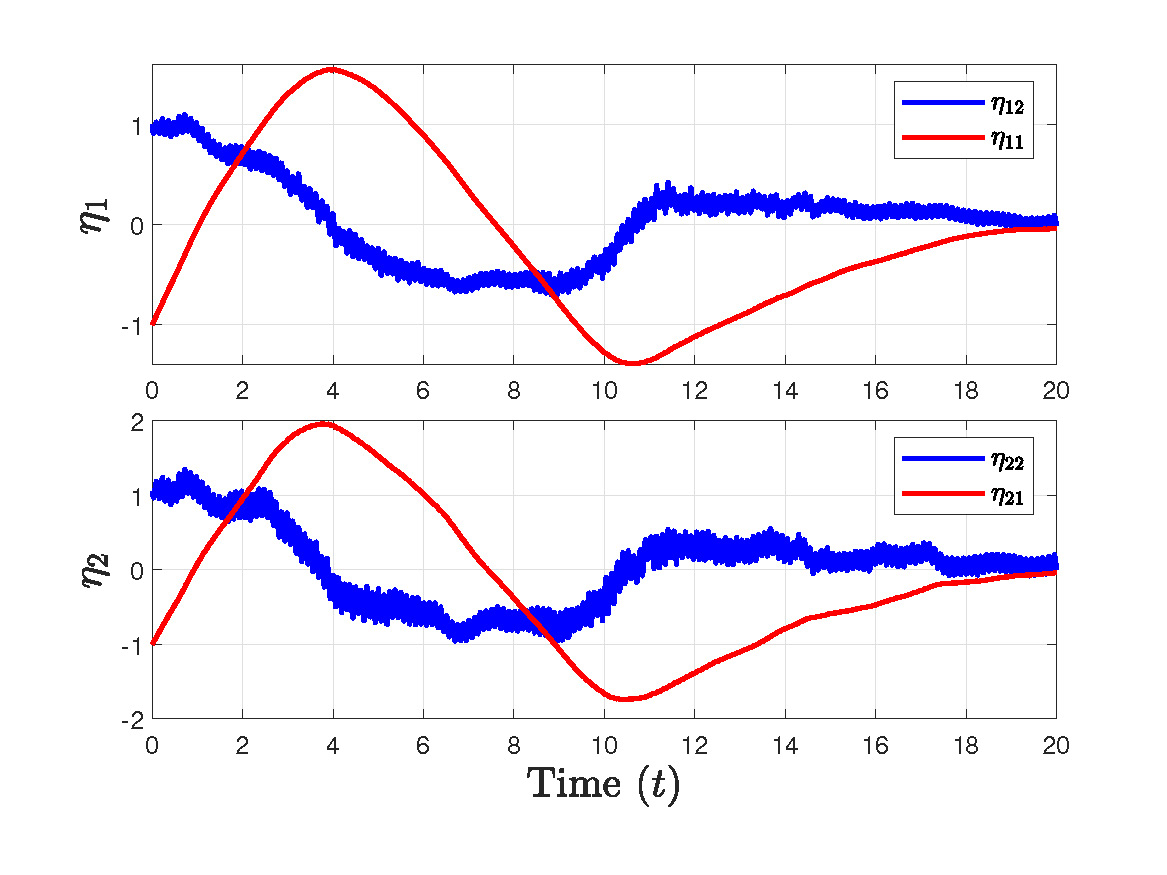}}}
\end{picture}
\end{center}
\caption{Tracking errors under the RR protocol case and the ETMs \eqref{eqn-33}, where $T_{1}=T_{2}=0.01$ and $\Delta_{1}=\Delta_{2}=0.0015$.}
\label{fig-1}
\end{figure}

Here, we consider the case that the communication between the controller and the plant is via the ETMs and two communication networks and quantizers. The controller is applied via the ZOH devices and the networks are assumed to have $\ell_{i}=3$ nodes for $\mathfrak{q}^{i1}_{\p}$, $\mathfrak{q}^{i2}_{\p}$ and $u_{i}$, respectively. Set $\max_{i\in\{1, 2\}, j\in\{1, 2, 3\}}\mathfrak{n}^{i}_{j}=0.8$ and $\max_{i\in\{1, 2\}, j\in\{1, 2, 3\}}\Omega^{i}_{j}=0.6$. Hence, the applied feedback controller is given by $u^{i}_{\ct}=-c^{-1}_{i}[a_{i}(\sin(\hat{\mathfrak{q}}^{i1}_{\p})-\sin(\hat{\mathfrak{q}}^{i1}_{\rf})) -(\hat{\mathfrak{q}}^{i1}_{\p}-\hat{\mathfrak{q}}^{i1}_{\rf})+(\hat{\mathfrak{q}}^{i2}_{\p}-\hat{\mathfrak{q}}^{i2}_{\rf})]$. $u^{i}_{\f}$ is assumed to be transmitted to \eqref{eqn-32} directly, and $\hat{\mathfrak{q}}^{i1}_{\rf}, \hat{\mathfrak{q}}^{i2}_{\rf}$ are implemented in the ZOH fashion. That is, $u^{i}_{\ct}$ knows but does not depend on $\mathfrak{q}^{i1}_{\rf}, \mathfrak{q}^{i2}_{\rf}$.

Based on \eqref{eqn-31}-\eqref{eqn-32}, we obtain that $F_{\eta}=(F^{1}_{\eta}, F^{2}_{\eta})$ with $F^{i}_{\eta}=(\eta_{i2}, -a_{i}[\sin(\eta_{i1}
+\mathfrak{q}^{i1}_{\rf})-\sin(\mathfrak{q}^{i1}_{\rf})-\sin(\eta_{i1}+\mathfrak{q}^{i1}_{\rf}+e^{i1}_{\eta}+e^{i1}_{\rf})+\sin(\mathfrak{q}^{i1}_{\rf}
+e^{i1}_{\rf})]-(\eta_{i1}+e^{i1}_{\eta})-(\eta_{i2}+e^{i2}_{\eta})+\sum_{j=1, 2}b_{ij}(\eta_{1j}-\eta_{2j})+c_{i}e^{i}_{\f}+c_{i}e^{i}_{\ct})$, $F_{\rf}=(F^{1}_{\rf}, F^{2}_{\rf})$ with $F^{i}_{\rf}=(\mathfrak{q}^{i2}_{\rf}, -a_{i}\sin\mathfrak{q}^{i1}_{\rf}+\sum_{j=1, 2}b_{ij}(\mathfrak{q}^{1j}_{\rf}-\mathfrak{q}^{2j}_{\rf})+c_{i}u^{i}_{\f})$, $G_{\ax}=(-F_{\eta}, 0)$, $G_{\rf}=-F_{\rf}$ and $G_{\f}=-(\dot{u}^{1}_{\f}, \dot{u}^{2}_{\f})$. In addition, $|G^{i}_{\ax}|\leq D_{i}|e_{i}|+|\eta_{i2}|+|(b_{i1}-1)\eta_{i1}+(b_{i2}-1)\eta_{i2}|+|b_{i1}\eta_{(3-i)1}|+|b_{i2}\eta_{(3-i)2}|
+2a_{i}|e^{i}_{r}|+c_{i}|e^{i}_{\f}|$ with $D_{i}=\sqrt{3}\max\{1+a_{i}, c_{i}\}$. From \cite{Ren2019tracking}, we choose the appropriate Lyapunov function $W_{i}(e_{i}, \mu_{i}, m_{i}, \kappa_{i}, \tau_{i}, b_{i})$. For instance, $W_{i}(e_{i}, \mu_{i}, m_{i}, \kappa_{i}, \tau_{i}, b_{i}):=\omega_{i}|e^{i}_{\ax}|+|\mu_{i}|$ for the TOD protocol, where $\omega_{i}\in(0, (1-\max_{j}\Omega^{i}_{j})/\max_{j}\mathfrak{n}^{i}_{j})$. $|\partial W(e_{i}, \mu_{i}, m_{i}, \kappa_{i}, \tau_{i}, b_{i})/\partial e_{i}|\leq M_{i}$ with $M_{i}=\sqrt{\ell_{i}}$ for the RR protocol case and $M_{i}=1$ for the TOD protocol case. Assumption \ref{asn-5} holds with $\lambda_{i}=\max\{\sqrt{(\ell_{i}-1)/\ell_{i}}, \omega_{i}\mathfrak{m}_{i}\mathfrak{n}^{i}_{j}+\Omega^{i}_{j}\}$ and $\alpha_{3i}=\alpha_{4i}=\alpha_{5i}=\alpha_{6i}=0$. Assumption \ref{asn-6} holds with $L_{i0}=M_{i}D_{i}, L_{i1}=M^{2}_{i}D_{i}/\lambda_{i}$, $H_{i0}(x, e)=H_{i1}(x, e)=M_{i}(|\eta_{i2}|+|(b_{i1}-1)\eta_{i1}+(b_{i2}-1)\eta_{i2}|+|b_{i1}\eta_{(3-i)1}|+|b_{i2}\eta_{(3-i)2}|)$, $\sigma_{1i0}(v)=\sigma_{1i1}(v)=c_{i}M_{i}v$ and $\sigma_{2i0}(v)=\sigma_{2i1}(v)=2a_{i}M_{i}v$ for $v\geq0$.

To verify Assumption \ref{asn-7}, define $V(\eta):=\sum^{2}_{i=1}\phi_{i1}\eta^{2}_{i1}+\phi_{i2}\eta_{i1}\eta_{i2}+\phi_{i3}\eta^{2}_{i2}$, where $\phi_{i1}, \phi_{i2}, \phi_{i3}$ are chosen to make $V$ satisfy \eqref{eqn-22}. Assume that there exist time-varying parameters $\hat{a}_{i}, \tilde{a}_{i}\in[-a_{i}, a_{i}]$ such that $a_{i}[\sin(\eta_{i1}+\mathfrak{q}^{i1}_{\rf})-\sin(\eta_{i1}+\mathfrak{q}^{i1}_{\rf}+e^{i1}_{\eta}+e^{i1}_{\rf})]
=\hat{a}_{i}(e^{i1}_{\eta}+e^{i1}_{\rf})$ and $a_{i}[\sin(\mathfrak{q}^{i1}_{\rf})-\sin(\mathfrak{q}^{i1}_{\rf}+e^{i1}_{\rf})]=\tilde{a}_{i}e^{i1}_{\rf}$. Thus, using twice the fact that $2\mathfrak{ab}\leq\mathfrak{ca}^{2}+\mathfrak{b}^{2}/\mathfrak{c}$ for all $\mathfrak{a}, \mathfrak{b}\geq0$ and $\mathfrak{c}>0$, we get that $\langle\nabla V(\eta), F_{\eta}(\delta, x, e, \mu)\rangle\leq\sum^{2}_{i=1}[-\phi_{i1}\eta^{2}_{i1}+(2\phi_{i1}-2\phi_{i3}
-\phi_{i2})\eta_{i1}\eta_{i2}-(2\phi_{i3}-\phi_{i1})\eta^{2}_{i2}+(\phi_{i2}\eta_{i1}+2\phi_{i3}\eta_{i2})(b_{i1}(\eta_{11}-\eta_{21})
+b_{i2}(\eta_{12}-\eta_{22}))+0.5(\varrho^{-1}_{i0}+\varrho^{-1}_{i1})(\phi_{i1}\eta_{i1}+2\phi_{i3}\eta_{i2})^{2}
+0.5\varrho_{i0}D_{i}|e_{i}|^2+\varrho_{i1}(4a^{2}_{i}|e^{i}_{\rf}|^{2}+c^{2}_{i}|e^{i}_{\f}|^{2})]$, where $\varrho_{i0}, \varrho_{i1}>0$ are defined in \eqref{eqn-28}. Therefore, if $\phi_{1}, \phi_{2}, \phi_{3}$ are chosen such that \eqref{eqn-22} holds and $-H^{2}_{ib_{i}}(x, e)-K_{ib_{i}}(x, e, \mu, m)-\varphi_{ib_{i}}(z_{i})\geq-\phi_{i1}\eta^{2}_{i1}+(2\phi_{i1}-2\phi_{i3}-\phi_{i2})\eta_{i1}\eta_{i2}-(2\phi_{i3}-\phi_{i1})\eta^{2}_{i2}
+(\phi_{i1}\eta_{i1}+2\phi_{i3}\eta_{i2})(b_{i1}(\eta_{11}-\eta_{21})+b_{i2}(\eta_{12}-\eta_{22}))+0.5(\varrho^{-1}_{i0}
+\varrho^{-1}_{i1})(\phi_{i1}\eta_{i1}+2\phi_{i3}\eta_{i2})^{2}$, then Assumption \ref{asn-7} is verified with $\theta_{ib_{i}}(v)=\pi_{i}v^{2}$, $\gamma_{i0}=\sqrt{\pi_{i}+\varrho_{i0}D^{2}_{i}}$, $\gamma_{i1}=\sqrt{\pi_{i}+\varrho_{i1}M_{i}D^{2}_{i}/\lambda^{2}_{i}}$, $\zeta_{1ib_{i}}(v)=\varrho_{i1}a^{2}_{i}|v|^{2}$, $\zeta_{4ib_{i}}(v)=\varrho_{i1}a^{2}_{i}|v|^{2}$ and $\pi_{i}>0$ is arbitrarily small.

\begin{figure}[!t]
\begin{center}
\begin{picture}(70,100)
\put(-70,-20){\resizebox{70mm}{40mm}{\includegraphics[width=2.5in]{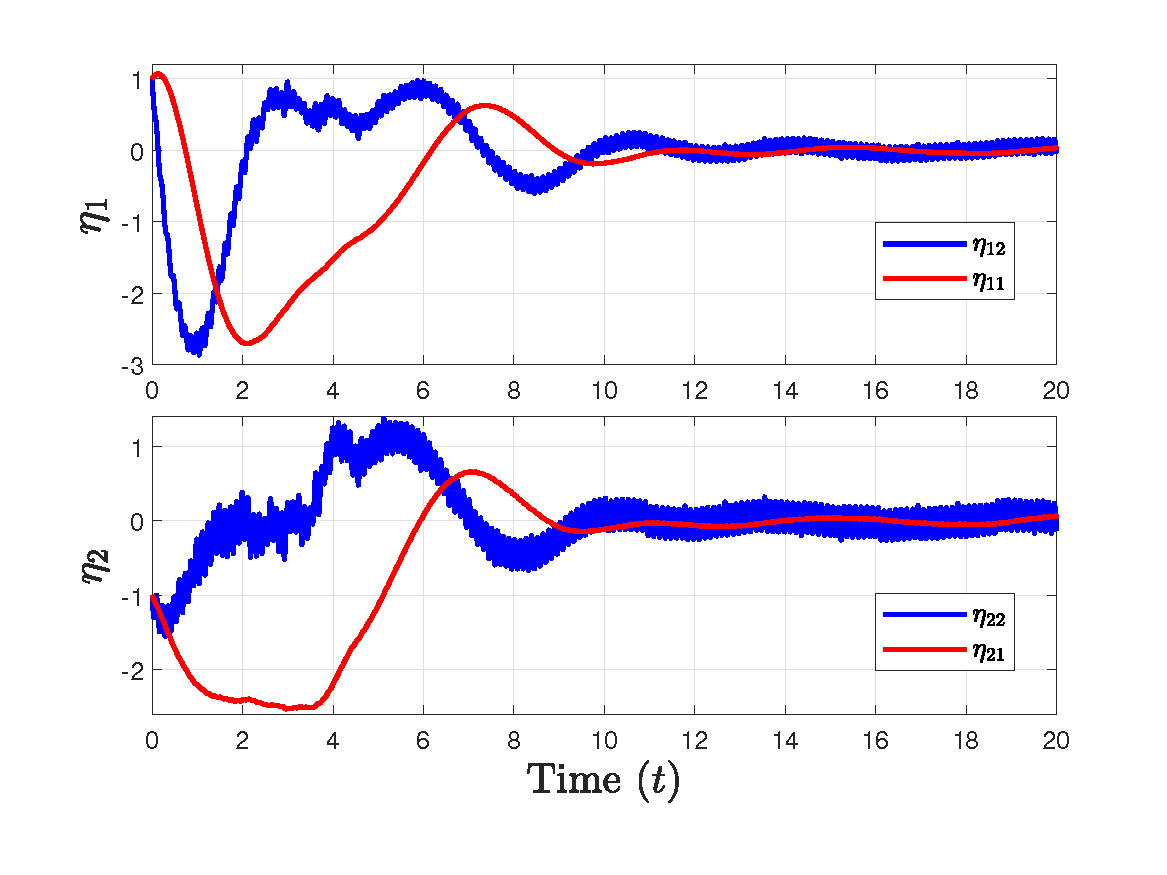}}}
\end{picture}
\end{center}
\caption{Tracking errors under the TOD protocol case and the ETMs \eqref{eqn-33}, where $T_{1}=T_{2}=0.014$ and $\Delta_{1}=\Delta_{2}=0.0025$.}
\label{fig-2}
\end{figure}

To satisfy the aforementioned conditions, we choose $\phi_{11}=8, \phi_{12}=12, \phi_{13}=6, \phi_{21}=5, \phi_{22}=7, \phi_{23}=9, a_{1}=9.81*0.2, a_{2}=9.81*0.3, c_{1}=2, c_{2}=4, \pi_{i}=0.005, \varrho_{i0}=0.05$ and $\varrho_{i1}=\varrho_{i0}M_{i}/\lambda_{1}$. Thus, $L_{10}=8.8860, L_{11}=18.8501, L_{20}=12, L_{21}=25.4558, \gamma_{10}=22.9436, \gamma_{11}=48.6706, \gamma_{20}=30.9839, \gamma_{21}=65.7267$ for the RR protocol case; $L_{10}=5.1303, L_{11}=10.8831, L_{20}=6.9282, L_{21}=14.6969, \gamma_{10}=22.9436, \gamma_{11}=28.1, \gamma_{20}=30.9839, \gamma_{21}=37.9473$ for the TOD protocol case. By the detailed computation, $\overbar{\rho}_{1}=0.0501$ and $\overbar{\rho}_{2}=0.0371$ for RR and TOD protocols. Hence, $\rho_{i}\in(0,\overbar{\rho}_{i})$, and the ETM is
\begin{align}
\label{eqn-33}
\Gamma_{i}=-b_{i}\gamma_{i}|(e^{i}_{\eta}, e^{i}_{\rf}, \mu_{i})|^{2}+(1-b_{i})\rho_{i}\bar{\lambda}_{i}|\eta_{i}|^{2}\geq0.
\end{align}
Set $\phi_{10}(0)=\phi_{11}(0)=1.1023$ and $\phi_{20}(0)=\phi_{21}(0)=0.8816$ for the RR protocol case, and we have $T_{1}=0.0256, \Delta_{1}=0.0064, T_{2}=0.0161$, and $\Delta_{2}=0.0026$. Set $\phi_{10}(0)=\phi_{11}(0)=\phi_{20}(0)=\phi_{21}(0)=1.0468$ for the TOD protocol case, and we have $T_{1}=0.0279, \Delta_{1}=0.00445, T_{2}=0.02115$, and $\Delta_{2}=0.0032$. To simplify the simulation, the transmission intervals and the transmission delays are constants. Under the ETM \eqref{eqn-28}, Figs. \ref{fig-1}-\ref{fig-2} show the convergence and boundedness of tracking errors in RR and TOD protocol cases, respectively.

The numbers of information transmission in different cases are presented in Table \ref{tab-1}. Note that we consider 23 units of time for the RR case and 32 units of time for the TOD case. Therefore, the transmission numbers are the same (2000 times) in the time-triggered case \cite{Postoyan2014tracking, Ren2019tracking}, whereas the transmission numbers are reduced to different extents in the event-triggered case. In particular, the transmission numbers of two networks in the RR case are less than these in the TOD case.

\section{Conclusions}
\label{sec-conclusion}

We presented a Lyapunov-based emulation approach for the event-triggered tracking control problem of NQCS, where the information communication is via multiple asynchronous networks. To deal with the considered problem, we proposed a new hybrid model, and then established sufficient conditions and designed decentralized event-triggered mechanisms. The tradeoff between the MASP and the MAD was determined to guarantee the tracking performance. The effectiveness of the proposed approach was illustrated via a numerical example.

\begin{table}[!t]
\centering
\caption{Comparison of transmission numbers in different triggering cases}
\label{tab-1}
\begin{tabular}{p{0.08\textwidth}>{\centering}p{0.08\textwidth}>{\centering}p{0.08\textwidth}>{\centering\arraybackslash}p{0.1\textwidth}}\hline
\multirow{2}{*}{Network}&\multicolumn{2}{c}{Event-triggering}&\multirow{2}{*}{Time-triggering\vspace{8pt}}\\\cline{2-3}
&RR case &TOD case& \cite{Postoyan2014tracking, Ren2019tracking}  \\
\hline
Network 1 &887&1606&2000\\
Network 2&1448&1838&2000\\
\hline
\end{tabular}
\end{table}

\bibliographystyle{IEEEtran}

\end{document}